\documentclass[doublecol]{epl2} 

\title{How does degree heterogeneity affect an order-disorder transition?}
\shorttitle{How does degree heterogeneity affect an order-disorder transition?} 

\author{R. Lambiotte\inst{1}}
\shortauthor{R. Lambiotte}

\institute{                    
  \inst{1} GRAPES, Universit\'e de Li\`ege, Sart-Tilman, B-4000 Li\`ege, Belgium
}


\abstract{
We focus on the role played by the node degree distribution on the way collective phenomena emerge on complex networks. To address this question, we focus analytically on a typical model for cooperative behaviour, the Majority Rule, applied to dichotomous networks. The latter are composed of two kinds of nodes, each kind $i$ being characterized by a degree $k_i$. Dichotomous networks are therefore a simple instance of heterogeneous networks, especially adapted in order to reveal the effect of degree heterogeneity. Our main result are that degree heterogeneity affects the location of the order-disorder transition and that the system exhibits non-equipartition of the average opinion between the two kinds of nodes. This effect is observed in the ordered phase and in the disordered phase.
}

\begin{document}

\maketitle

\section{Introduction}

It is more and more common, nowadays, to use models and tools from Statistical Physics in order to describe the emergence of collective phenomena in social systems. Amongst other examples, one may think of opinion formation \cite{galam,szn}, rumour or disease spreading \cite{watts2002,boguna}, language dynamics \cite{dall}, etc. In these models, agents are located at the node of a graph and are endowed with a finite number of available states, e.g. two states - spin up and spin down. The links between nodes represent the relations (e.g. friendship, co-authorship) between agents. Most of the models are based on local {\em attractive} interactions, i.e. agents copy the behaviour of their direct neighbours, while random changes also take place, thereby mimicking the effect of individual choices. Contrary to classical problems of Statistical Physics, though, the underlying network structure is not a $d$-dimensional regular lattice, but a complex network \cite{bara}, whose node degree (number of links per node) and other internal properties may vary from one node to another.

Several works have revealed that a given model may exhibit very different (even qualitatively) behaviours depending on its underlying topology \cite{boguna}. Important factors are for instance the degree heterogeneity \cite{sood}, the presence of communities \cite{lambiotte,lam2} or the small-world property \cite{bar}.
From a practical point of view, it is therefore important to elucidate how these structural properties affect critical behaviour if one intends to reproduce the emergence of collective phenomena as they take place in realistic situations.  A typical example would be the propagation of rumours in a social network, which is of primordial interest for viral marketing issues \cite{viral}. From a theoretical point of view, it is challenging to develop tools and methods in order to determine the influence of the network topology on the way the system orders (or not). Such theoretical studies have been performed in some specific cases, e.g. the Voter Model \cite{sood} or the Unanimity Model \cite{lam0}, but a coherent and unifying framework is still needed.

In this Letter, we address this problem by focusing on a variant of the 
majority rule model (MR), that is a typical model for consensus formation \cite{galam}. Its simple ingredients allow a very complete analytical description in the mean-field approximation \cite{krap}. One should also stress that, contrary to the voter model  \cite{liggett}, MR does not conserve average magnetization \cite{redner}. In the following, we focus on a variant of MR that includes random flips ({\em thermal fluctuations}) and study analytically the effect of the degree heterogeneity on the phase diagram. To do so, we introduce {\em dichotomous networks}, in which there are two sorts of nodes, each sort being characterized by a degree $k_1$ or $k_2$. It is shown that the system undergoes a transition from a disordered phase to an ordered phase for weak enough random effects ($\sim$ low temperature). Our main results are that the location of this transition depends on the degree heterogeneity $\gamma \equiv k_2/k_1$. Moreover, the system is shown to exhibit non-equipartition of the average {\em magnetization} (each sort of nodes is characterized by a different average opinion/spin) when $\gamma \neq 1$. This is observed in the ordered and in the disordered phase.

\section{Majority Rule}

The network is composed of $N$ nodes, each of them endowed with an opinion $o_i$ that can be $\alpha$ or $\beta$. At each time step, one of the nodes is randomly selected. Two processes may take place.  i) With probability $q$, the selected node $s$ randomly changes its opinion: 
\begin{eqnarray}
o_s \rightarrow \alpha  & {\rm with ~probability~} 1/2,\cr
o_s \rightarrow \beta  & {\rm with ~probability~}1/2.
\end{eqnarray}
ii) With probability $1-q$, two neighbouring nodes of $s$ are also selected and the three agents in this {\em majority triplet} all adopt the state of the local majority. The parameter $q$ therefore measures the competition between individual choices, that have a tendency to randomize the system, and neighbouring interactions, that tend to homogenize the opinion of agents. In the case $q=0$, it is well-known that the system asymptotically reaches global consensus where all nodes share the same opinion \cite{krap}. In the other limiting case $q=1$, the system is purely random and the average (over the realizations of the random process) number of nodes with opinion $\alpha$ at time $t$, denoted by $A_t$, goes to $N/2$ for $t$ large. 

\section{Homogeneous network}

In this section, we assume that the network of individuals is highly connected and homogeneous, i.e. all the nodes have the same degree. In that case, the mean-field rate equation for $A_t$ reads 
\begin{eqnarray}
\label{simple}
A_{t+1} = A_t +  q (\frac{1}{2} - a ) - 3 (1-q) a (1 - 3 a + 2 a^2),
\end{eqnarray}
where $a_t=A_t/N$ is the average proportion of nodes with opinion $\alpha$.
The term proportional to $q$ accounts for the random flips and the last term for local majorities. This comes from the fact that the probability for two nodes $\alpha$ ($\beta$) and one node $\beta$ ($\alpha$) to be selected is $3 a^2 (1-a)$ ($3 a (1-a)^2$), so that the total contribution to the evolution of $A_t$  is
\begin{eqnarray}
\label{w}
 W = 3 \left( a^2 (1-a)-a (1-a)^2 \right) = - 3 a (1 - 3 a + 2 a^2).
 \end{eqnarray} 
 Let us stress that Eq. (\ref{w}) differs from Eq. (2) of \cite{lam0} by a factor 3. In \cite{lam0}, this factor could be absorbed in the time scale as it did not play a relevant role.
It is straightforward to show that $a=1/2$ is always a stationary solution of Eq.~(\ref{simple}), as expected from symmetry reasons. This is obvious after rewriting the evolution equation for the quantities $\Delta=A-N/2$ and $\delta=a-1/2$
\begin{eqnarray}
\label{simpleSim}
\Delta_{t+1} = \Delta_t + \frac{\delta}{2} \left(3-5q - 12 (1-q) \delta^2 \right),
\end{eqnarray}
from which one finds that the symmetric solution $a=1/2$ ceases to be stable when $q<3/5$, and that the system reaches the following asymmetric solutions in that case
\begin{eqnarray}
\label{solution}
a_- = \frac{1}{2} - \sqrt{\frac{3 - 5 q}{12(1-q)}}, ~~
a_+ = \frac{1}{2} +  \sqrt{\frac{3 - 5 q}{12(1-q)}}.
\end{eqnarray}
The system therefore undergoes an order-disorder transition at $q=3/5$. Under this value, a collective opinion has emerged due to the {\em imitation} between neighbouring nodes. Let us stress that Eqs. (\ref{solution})
 respectively converge  to $a_-=0$ and $a_+=1$ in the limit $q \rightarrow 0$.

\section{Degree heterogeneity}

The main goal of this Letter is to understand how the degree distribution of the underlying network affects collective behaviours as those of the previous section. Contrary to more phenomenological approaches, where the effects of the degree heterogeneity are brought to light by comparing the behaviour of a model on several kinds of networks \cite{suchecki} (e.g. Erd\"os-Renyi, scale-free, etc), we prefer to address the problem from an analytical and more fundamental point of view. To do so, we generalize homogeneous networks in the most 
natural way by considering random networks whose nodes may be divided into two classes, the nodes in different classes being characterized by a different degree, $k_1$ or $k_2$. This binary mixture, that we call a {\em dichotomous network} is particularly suitable in order to reveal the role of degree distribution. Indeed, the degree heterogeneity is tunable through the parameter $\gamma=k_2/k_1$. When $\gamma \rightarrow 1$, one recovers an homogeneous network. Let us also emphasize that dichotomous networks differ from usual bipartite networks \cite{bipartite} by the fact that links may exist between nodes of the same class.

The underlying topology is therefore determined by the values $k_1$ and $k_2$, and by $N_{k_1}$ and $N_{k_2}$ that are the number of nodes of each class. In the following, we will assume that $N_{k_1}=N_{k_2}$ for the sake of simplicity. A more complete analysis for the full range of parameters will be considered elsewhere.  We are interested in $A_{1;t}$ and $A_{2;t}$ that are the average number of nodes $1$ and $2$ with opinion $\alpha$. By construction, each  node is selected with the same probability during one time step, but those with a higher degree have a larger probability to be among the neighbours of the selected node, i.e. to be in the majority triplet. This effect will have to be taken into account in order to generalize Eq. (\ref{simple}).

By construction, the probability that the selected node has a degree $k_i$ is $1/2$, but the probability that the neighbour of this selected node has a degree $k_j$  is $ k_j/(k_1+k_2)$ (one assumes that there are no correlations between the degrees of neighbouring nodes). Consequently, the probability that the selected node has a degree $k_1$ and that both of its selected neighbours have a degree $k_2$ is 
\begin{eqnarray}
 \frac{k_2^2}{2 (k_1 + k_2)^2}= \frac{\gamma^2}{2 (1 + \gamma)^2}.
\end{eqnarray}
Similarly, the probability that the selected node has a degree $k_1$, that one of its neighbours has a degree $k_1$ and that the other neighbour has a degree $k_2$ is
\begin{eqnarray}
 \frac{k_1 k_2}{ (k_1 + k_2)^2}= \frac{\gamma}{ (1 + \gamma)^2},
\end{eqnarray}
while the probability that all three nodes have a degree $k_1$ is
\begin{eqnarray}
 \frac{k_1^2}{2 (k_1 + k_2)^2}= \frac{1}{2 (1 + \gamma)^2}.
\end{eqnarray}
The sum of these three probabilities is normalized and the probabilities of events when the selected node has a degree $k_2$ are found in the same way. Putting all contributions together, one finds the probabilities $P_{(x,y)}$ that $x$ nodes 1 and $y$ nodes 2 belong to the majority triplet
\begin{eqnarray}
\label{heterog}
P_{(3,0)} =   \frac{ 1}{2 (1+\gamma)^2}, ~~~~
P_{(2,1)} =   \frac{ 1 + 2 \gamma}{2 (1+\gamma)^2}, \cr
P_{(1,2)} =  \frac{ 2 \gamma + \gamma^2}{2 (1+\gamma)^2}, ~~~~
P_{(0,3)} = \frac{  \gamma^2}{2 (1+\gamma)^2},
\end{eqnarray}
where the normalization $\sum_{xy} P_{(x,y)}=1$ is verified. In order to derive the coupled equations generalizing Eq. (\ref{simple}) for $A_{i;t}$, one needs to evaluate the possible majority processes taking place when a triplet $(x,y)$ is selected. Let us focus on the case $(2,1)$ as an example. In that case, the number of nodes $A_{1}$, $A_{2}$  will change due to the contributions
\begin{eqnarray}
\label{coco}
 W_{1,(2,1)} &=&  [2 a_1 a_2 (1-a_1)- 2 a_1 (1-a_2) (1-a_1)]  \cr
 W_{2,(2,1)} &=&  [a_1^2 (1-a_2)-  a_2 (1-a_1)^2], 
 \end{eqnarray}
  where $a_i$ is the proportion of nodes with opinion $\alpha$ in the class $i$.
The first line accounts for cases when one node 1 and one node 2 have the same opinion but disagree with a node  1, while the second line accounts for cases when the 2 nodes  1 have the same opinion but disagree with the node  2. The other situations $(x,y)$ are treated similarly \cite{lambi}. Putting all the contributions together, one arrives at the set of non-linear equations
 \begin{eqnarray}
 \label{complicated}
  A_{1;t+1} - A_{1;t}&=& \frac{q}{4} - \frac{q a_1}{2} + \frac{(1-q)}{2 (1+\gamma)^2} [ 3 (a_1^2 b_1- a_1 b_1^2) \cr
  &+& 2 (1 + 2 \gamma) (a_2 a_1 b_1- a_1 b_2 b_1) \cr
  &+&  (2 \gamma + \gamma^2) (a^2_2 b_1- a_1 b^2_2)] \cr
   A_{2;t+1} - A_{2;t}&=& \frac{q}{4} - \frac{q a_2}{2} + \frac{(1-q)}{2 (1+\gamma)^2} [3 \gamma^2 (a_2^2 b_2- a_2 b_2^2)\cr
   &+& 2 (2 \gamma + \gamma^2) (a_1 a_2 b_2- a_2 b_1 b_2) \cr
   &+&  (1 + 2 \gamma) (a^2_1 b_2- a_2 b^2_1)], 
 \end{eqnarray} 
  where  $b_i$ is the proportion of nodes with opinion  $\beta$ in the class $i$ ($b_i=1-a_i$). One verifies by summing the equations for $A_1$ and $A_2$ that  Eq. (\ref{simple}) is recovered in the limit $\gamma=1$ (with $a=(a_1+a_2)/2$), as expected due to the indistinguishability of the nodes in that case.
  
\begin{figure}

\onefigure[angle=-90,width=3.45in]{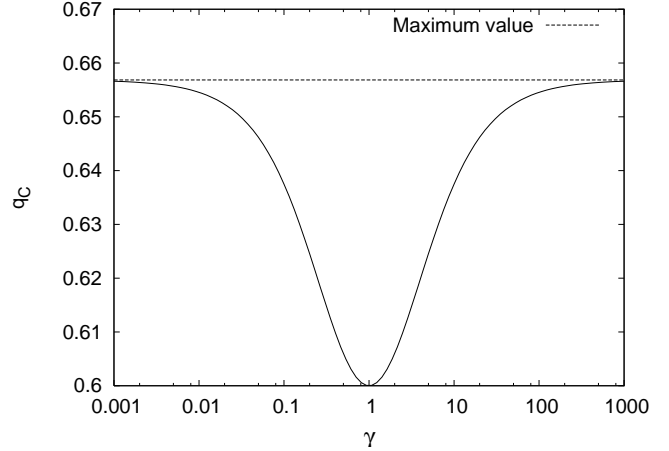}

\caption{Location of the order-disorder transition $q_c$ as a function of $\gamma$. The x-scale is logarithmic in order to make the symmetry $\gamma \leftrightarrow  \gamma^{-1}$ clear.}
\label{fig1}
\end{figure}

 It is easy to show that $a_1=a_2=1/2$ is always a stationary solution of the above set of equations. The stability analysis is performed by looking at deviations to this solution $a_1= 1/2 + \delta_1$, $a_2=1/2+\delta_2$ and keeping only linear corrections. In the continuous time limit and after re-scaling the time units, the evolution equations for these deviations read
 \begin{eqnarray}
 \label{linearised}
 \partial_t \delta_1 &=& - \frac{-3 + 5 q + 2 \gamma (1+q) + \gamma^2 (1+q)}{4 (1+\gamma)^2} \delta_1 \cr
&+& \frac{(1+ 4 \gamma + \gamma^2) (1-q)}{2 (1+\gamma)^2} \delta_2\cr
 \partial_t \delta_2 &=& - \frac{1 +  q + 2 \gamma (1+q) + \gamma^2 (-3+5 q)}{4 (1+\gamma)^2} \delta_2 \cr
&+& \frac{(1+ 4 \gamma + \gamma^2) (1-q)}{2 (1+\gamma)^2} \delta_1.
 \end{eqnarray} 
 The stability analysis requires to look at the eigenvalues of the above evolution matrix and to find the critical value of $q$ at which the real part of one eigenvalue becomes positive. Lengthy calculations lead to
 \begin{eqnarray}
 \label{qc}
q_c(\gamma)= \frac{8-16 \gamma +8 \gamma^2 + 16 \sqrt{K}}{24+16 \gamma + 24 \gamma^2 + 
16 \sqrt{K}}
 \end{eqnarray} 
 where $K=2 + 8 \gamma + 
16 \gamma^2 + 8 \gamma^3 + 2 \gamma^4$. In the limiting case $\gamma \rightarrow 1$, one recovers the known result $q_c=3/5$. It is also possible to verify (see Fig. 1) that this relation is symmetric under the changes $\gamma \leftrightarrow  \gamma^{-1}$, i.e. under an exchange of nodes 1 and 2. Moreover, the maximum value is obtained for $\gamma \rightarrow 0$ and $\gamma \rightarrow \infty$, $q_c(0)=q_c(\infty)=(1+2 \sqrt{2})/(3+2 \sqrt{2})$. Our first conclusion is therefore that the location of the order-disorder transition depends in a non-trivial way on the degree heterogeneity $\gamma$, even though these deviations remain of small magnitude (maximum $10 \%$). Moreover, the location of the transition is shifted to higher values of $q$ when the system is more hereogeneous and $q_c(1) \leq  q_c(\alpha)$. 

It is important to remind that node degree heterogeneity leads to a heterogeneity of the probability to belong to a {\em majority triplet} (see Eqs. (\ref{heterog})) and therefore to a heterogeneity of the {\em frequency of interaction}. In contrast, the frequency of random flips remains the same for the two types of nodes. However, as emphasized above, the order-disorder transition comes from a competition between the frequency of random flips $\sim q$ and the frequency of majority triplets $\sim (1-q)$. It is therefore understandable that the heterogeneity of the system might lead to a shift of $q_c$. In order to highlight this effet in detail, it is interesting to consider the limiting case $\gamma=0$, where the network is so heterogeneous that any randomly chosen link arrives at a node of type 1. Consequently, a node $2$ has a vanishing probability to be chosen as a neighbour of the selected node and it may belong to a majority triplet only if it is the selected node itself. When $\gamma=0$, the equations of evolution for the quantities $A_i$ read
 \begin{eqnarray*}
A_{1;t+1} - A_{1;t} &=& \frac{q}{4} - \frac{q a_1}{2} + \frac{(1-q)}{2 } [ 3 (a_1^2 b_1- a_1 b_1^2)\cr
&+& 2  (a_2 a_1 b_1- a_1 b_2 b_1)] \cr
A_{2;t+1} - A_{2;t}&=& \frac{q}{4} - \frac{q a_2}{2} + \frac{(1-q)}{2} [(a^2_1 b_2- a_2 b^2_1)] 
 \end{eqnarray*} 
 and it is straightforward, but lengthy, to show that the stationary solutions are either $
a_1=a_2=1/2$ or $a_1=1/2 \pm \delta_1, a_1=1/2 \pm \delta_2$ with
 \begin{eqnarray}
 \label{gamma0}
 \delta_1 &=& \frac{\sqrt{2 ( q^2 -1)  +  (1-q) \sqrt{25 - 22 q + q^2}} }{ \sqrt{12} (1-q)}\cr
\delta_2 &=& \frac{\delta_1}{2 (\delta_1^2 + \frac{1+q}{4(1-q)}) }.
 \end{eqnarray} 
 It is also possible to show that the ordered solutions exist only for $q<q_c(0)$, where $q_c(0)$ is given by (\ref{qc}). It is instructive to look at the linearized dynamics of $\delta_i$ around the disordered solution
  \begin{eqnarray}
 \partial_t \delta_1 &=& (3 - 5 q) \delta_1 + 2 (1-q) \delta_2\cr
 \partial_t \delta_2 &=&  2 (1-q) \delta_1  - (1 +  q ) \delta_2,
 \end{eqnarray} 
 whose matrix of evolution $L_{ij}$ is obviously asymmetric under the change $1 \leftrightarrow 2$, contrary to the case $\gamma=1$, where the linearized equations read
   \begin{eqnarray}
 \partial_t \delta_1 &=& - 4 q  \delta_1 + 6 (1-q)  \delta_2\cr
 \partial_t \delta_2 &=&  6 (1-q)  \delta_1  - 4 q  \delta_2.
 \end{eqnarray} 
These two sets of equations not only differ by their symmetry, but also by the relative importance of the diagonal terms $L_{ii}$ as compared to the cross terms $L_{ij}$ $i\neq j$ (such terms measure the coupling between the two kinds of nodes). This difference makes possible a shift of the value of $q_c$ where the determinant of $L_{ij}$ vanishes, i.e. where one eigenvalue vanishes. 
 
\begin{figure}

\onefigure[angle=-90,width=3.45in]{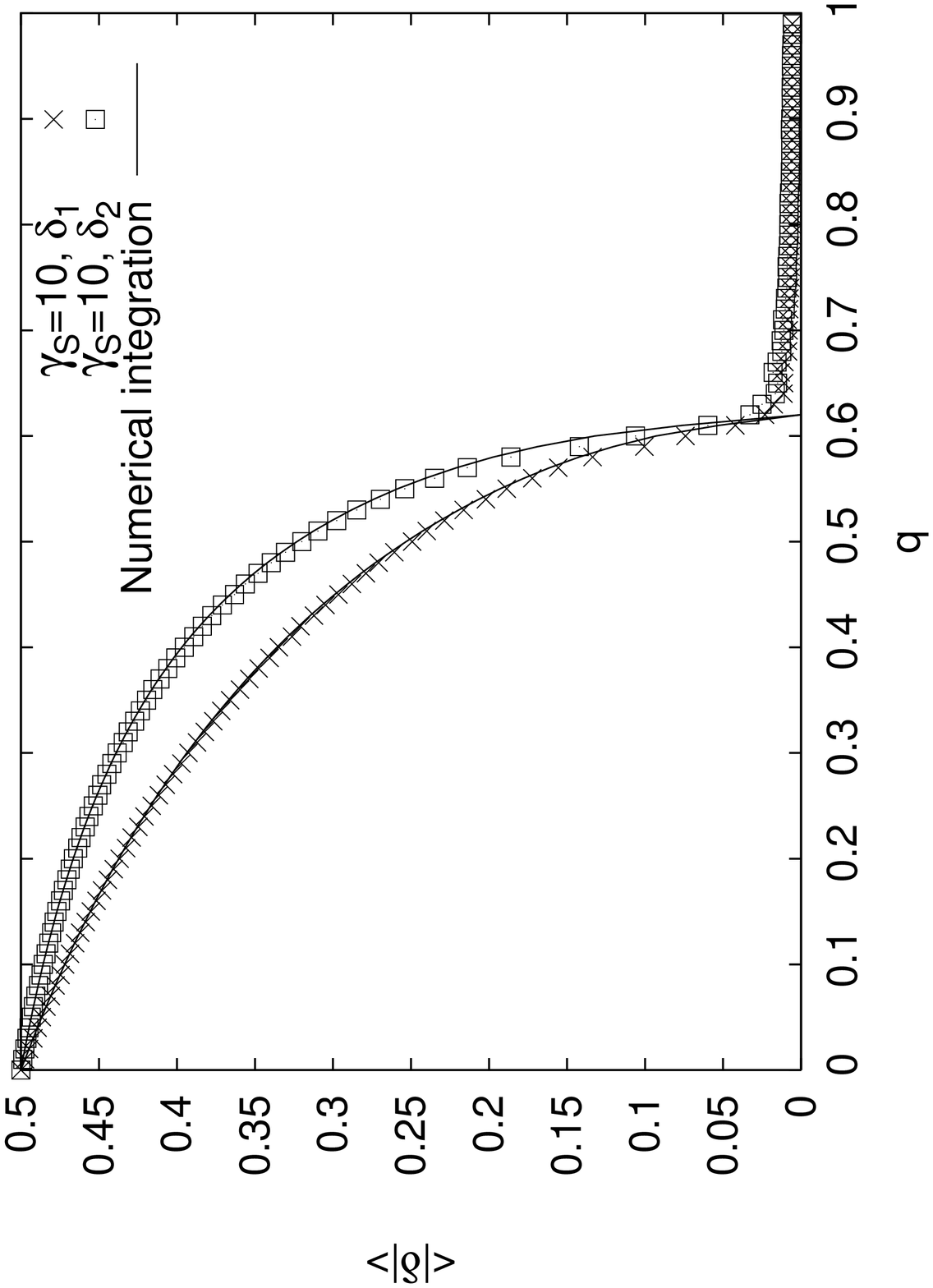}

\onefigure[angle=-90,width=3.45in]{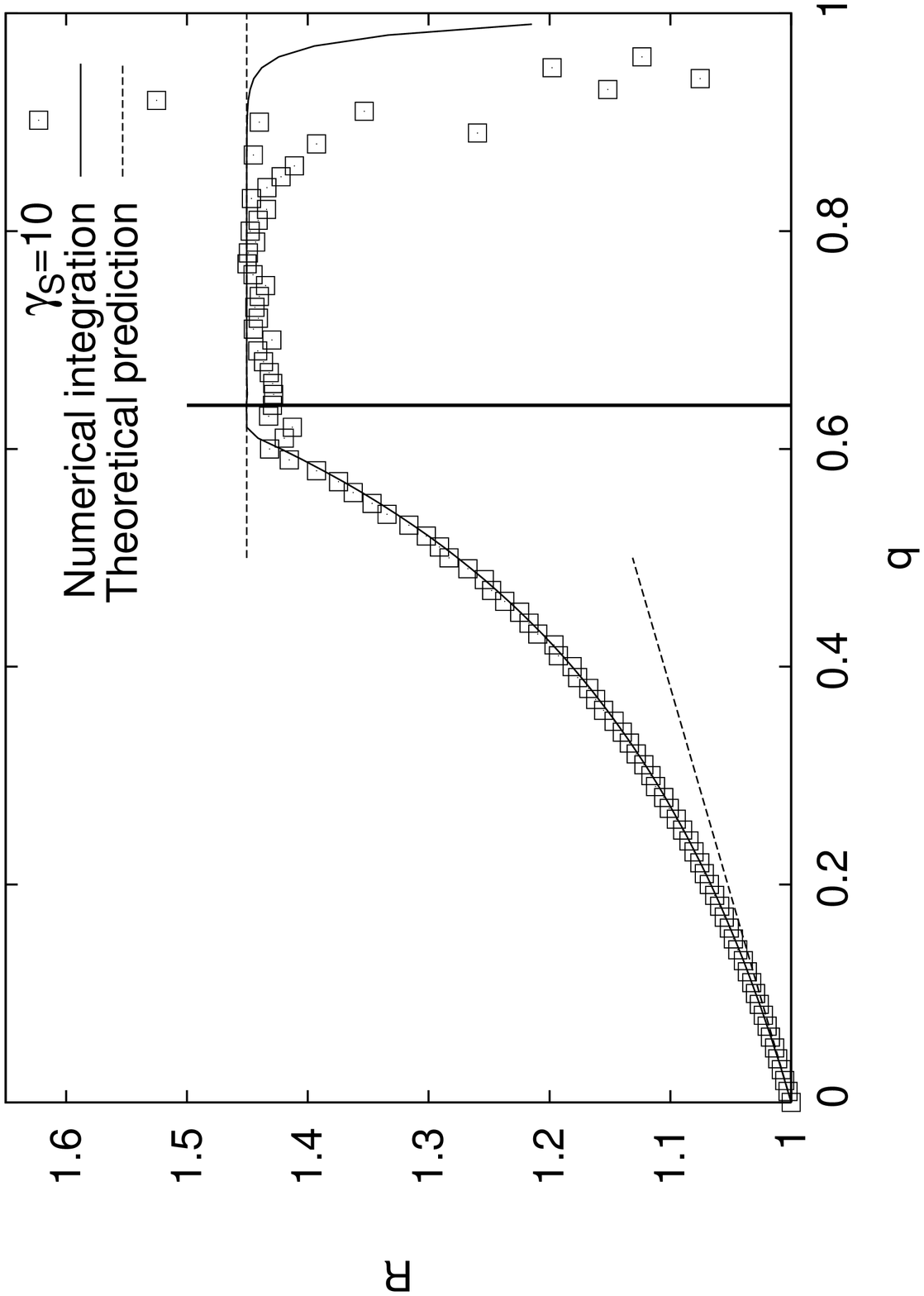}

\caption{Bifurcation diagram of $\delta_i(q)$ (upper figure) and $R(q)$ (lower figure) for $\gamma=3.162$ ($N=10^4$ nodes,  $p_1=0.005$, $\gamma_S=10$, see section {\bf Simulation results}). Solid lines are the asymptotic solutions obtained by integrating numerically Eqs. (\ref{complicated}). Simulation details are found in the main text. In the lower figure, the dashed line correspond to the theoretical predictions Eq. (\ref{appro}) and Eq. (\ref{delta}). The vertical line indicates the value of $q$ above which $\delta$ is measured during the relaxation to $\delta_i=0$, and not anymore in the asymptotic state $\delta_i\neq 0$.}
\label{fig2}
\end{figure}

Let us now return to our study of MR for general values of $\gamma$. In order to elucidate the behaviour of $a_1$ and $a_2$ below $q_c$, we have performed numerical integration of Eqs. (\ref{complicated}). It appears (see Fig. 2) that $a_1$ and $a_2$ reach different asymptotic values $a_{1,\infty}\neq a_{2,\infty}$ and that the class of nodes with the higher degree exhibit larger deviations to $1/2$ than the other class. This may be understood by the fact that nodes with a higher degree are more often selected in majority triplets, thereby triggering their tendency to reach consensus. In order to evaluate this non-equipartition of the average opinion between the two species of nodes, we introduce the ratio $R=\delta_2/\delta_1$. One observes (not shown in Fig. 2) that  $R=1$ when $\gamma=1$ and that increasing values of $\gamma$ lead to increasing values of $R$ at fixed $q$.

We tackle this problem from an analytical point of view by focusing on the limit $q \rightarrow 0$ and  looking for small deviations to global consensus $a_i=0+\epsilon_i$. By inserting this development into Eqs. (\ref{complicated}) and keeping linear terms, it is straightforward to show that the asymptotic values of $a_i$ are
$a_1= \frac{1+\gamma}{2 (5+\gamma)} q$ and $a_2= \frac{1+\gamma}{2 (1+5 \gamma)} q$,
from which one shows that

 \begin{eqnarray}
 \label{appro}
R \approx 1 + 4 \frac{ \gamma^2 - 1}{ (5+\gamma)(1+ 5 \gamma)} q. 
\end{eqnarray}
This solution is in perfect agreement with the numerical integration of Eqs. (\ref{complicated}) in the limit of small $q$, and is asymmetric, i.e. $R(\gamma^{-1})=R(\gamma)^{-1}$ (which leads to $R(\gamma^{-1})-1 = - (R(\gamma) - 1)$ in the linear approximation) under the change $\gamma \leftrightarrow  \gamma^{-1}$, as expected. One can also show that Eq. (\ref{appro}) with $\gamma=0$ is recovered from the exact solution (\ref{gamma0}), by dividing $\delta_2$ by $\delta_1$ and keeping only first corrections in $q$.

\begin{figure}
\onefigure[width=3.45in]{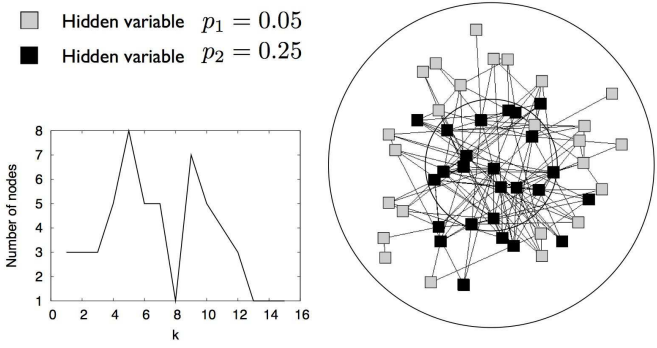}
\caption{Typical realization of a dichotomous network composed of $N=50$ nodes, with $p_1=0.05$ and $\gamma_S=5$. In the lower left corner, we plot the histogram of the node degrees that show two peaks. Each peak is associated to one class of nodes. The network is visualized thanks to the visone graphical tools \cite{visone} with a radial layout, i.e. the higher the degree of a node, the more central is its position in the graph. The hidden variable heterogeneity obviously leads to degree heterogeneity, while preserving its random structure.}
\label{fig3}
\end{figure}

Before going further, we would like to focus on a property that may appear intriguing at first sight. Namely, if one calculates the time evolution of $R$ from Eqs. (\ref{complicated}) above the critical point $q_c$, one finds that this quantity asymptotically converges to a constant value different from zero $R_{\infty} \neq 0$. In contrast, we have shown above that the deviations $\delta_1$ and $\delta_2$ go to zero for these values of $q$ ($a_i=1/2$ is a stable solution). Altogether, this suggests that the relaxation to the asymptotic state behaves like
$ \delta_1 = C_1 e^{-\lambda_1 t}$,
$ \delta_2 = C_2 e^{-\lambda_2 t}$,
where the relaxation rates are the same $\lambda_1=\lambda_2$ and where $C_2=R_{\infty} C_1$.
This behaviour reminds the non-equipartition of energy taking place in inelastic gases, for which it is well-known that different components of a mixture may exhibit different partial temperatures \cite{garzo5}. This analogy is clear after noting the similarities between the linearized equations (\ref{complicated}) and the equations for the partial temperatures $T_i$ \cite{lambi} obtained from Inelastic Maxwell Models \cite{ben1}.  

By using the same methods as in \cite{lambi}, one writes a closed  equation for $R$
 \begin{eqnarray}
 \partial_t R &=&   \frac{(1-q)}{2 (1+\gamma)^2} [(1+ 4 \gamma + \gamma^2) + 2 ( \gamma^2-1) R \cr
 &-& (1+ 4 \gamma + \gamma^2) R^2]
 \end{eqnarray} 
whose stationary solution is found to be
 \begin{eqnarray}
 \label{delta}
R = \frac{( \gamma^2-1) + \sqrt{2} \sqrt{1 + 4 \gamma + 8 \gamma^2 + 4 \gamma^3 + \gamma^4}}{(1+ 4 \gamma + \gamma^2)}.
 \end{eqnarray} 
This relation shows the same asymmetry as Eq. (\ref{appro}). One also verifies that $R$ goes to 1 when $\gamma \rightarrow 1$ and that $R$ goes to the finite values $(1+\sqrt{2})^{-1}$ and $(1+\sqrt{2})$ when $\gamma \rightarrow 0$ and $\gamma \rightarrow \infty$ respectively.

\section{Simulation results}

\begin{figure}

\onefigure[angle=-90,width=3.45in]{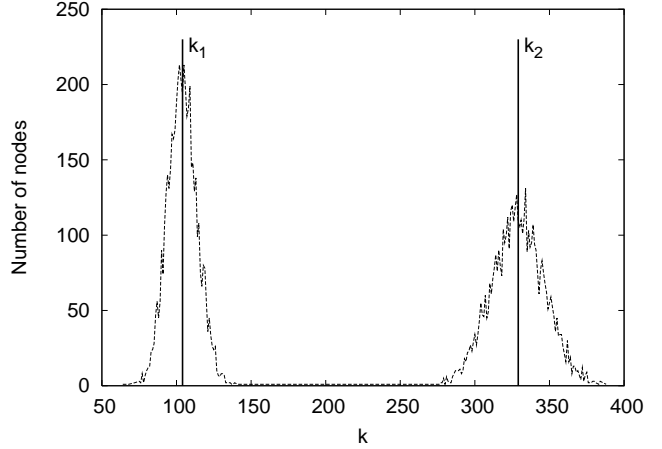}

\caption{Histogram of the node degrees for a dichotomous network composed of $N=10^4$ nodes, with $p_1=0.005$ and $\gamma_S=10$. The vertical lines indicate the average degrees $k_1\approx 104$ and $k_2 \approx 329$ obtained from Eqs. (\ref{vertical}). By construction, the surface under each {\em sub-distribution} is equal to $N/2=5 \times 10^3$.}
\label{fig4}
\end{figure}

By construction, the random steps of MR are easy to implement in a computer simulation. The only difficulty is to build the underlying dichotomous network, i.e. a network in which there are two characteristic node degrees but without internal correlations. We tackle this problem by  applying a method reminiscent of networks with {\em hidden variables} \cite{boguna2,fron,stefan} and {\em Superstatistics} \cite{beck,beck2,ausloos1}. Namely, we first prepare $N$ nodes and assign them hidden variables $h_i<1$, half of the nodes receiving the hidden variable $h_i=p_1$ and the other half the hidden variable $h_i=p_2$.
Next, each pair of nodes $(i, j)$ receives a link with a probability equal to $\sqrt{h_i h_j}<1$ (see Fig. 3).
Let us introduce $\gamma_S=p_2/p_1$ that measures the hidden variable heterogeneity. From the above definitions, one finds that the degree distributions of the nodes 1 and 2 are peaked around 
\begin{eqnarray}
\label{vertical}
k_{1} = \frac{N}{2} (p_1+\sqrt{p_1 p_2})\cr
k_{2} = \frac{N}{2} (p_2+\sqrt{p_1 p_2}).
 \end{eqnarray} 
The resulting network (see Fig. 4) is therefore a good candidate for dichotomous networks, where the effective $\gamma=k_2/k_1$ is related to $\gamma_S$ through the relation
\begin{eqnarray}
\gamma=\frac{p_2+\sqrt{p_1 p_2}}{p_1+\sqrt{p_1 p_2}} = \frac{\gamma_S+\sqrt{\gamma_S}}{1+\sqrt{\gamma_S}}.
 \end{eqnarray} 
Let us stress that the peaks of the degree distribution are not exactly delta functions, but their width is vanishingly small for large enough values of $k_1$ and $k_2$. 

The computer simulations presented in this Letter are performed with $N=10^4$ nodes, $p_1=0.005$ and $\gamma_S=10$, but other parameters have also been considered and lead to the same conclusions. The corresponding degree heterogeneity is therefore $\gamma \approx 3.162$. The stationary state of MR is measured by starting the simulation from the initial condition $a_i=1/2$ and letting the system relax toward its asymptotic state during 100 time steps/node. Afterwards, one measures $|\delta_1|$ and $|\delta_2|$ during 500 time steps/node, average over the time and over 100 realizations of the random process. This method gives an excellent agreement (see Fig. 2) with the numerical integration of Eqs. (\ref{complicated}) below the critical value $q_c$, but it is not applicable for measuring $R$ above this critical value. This is due to the fact that non-equipartition of opinion takes place during the relaxation to the asymptotic solution $a_i=1/2$ when $q>q_c$ and that this relaxation is indiscernible from the finite size fluctuations around $a_i=1/2$ in the long time limit. Consequently, we use another strategy in order to measure $R$ when $q>q_c$. Namely, the simulation is started from $a_i=0.7$, but $R$ is now evaluated during the relaxation, i.e. after 50 time steps/node, and averaged over 5000 realizations of the random process. These simulation results are in very good agreement with Eq. (\ref{delta}) (at least for $q<0.8$, above which fluctuations are very large) and confirm that $R$ does not depend on $q$ in the disordered phase. Finally, let us stress that the parameters used for the simulations imply that the degree of the nodes is very large (see Fig. 4). This choice ensures that the diameter of the network is very small and that the mean-field approximation used in order to derive Eqs. (\ref{complicated}) is  justifiable. It is therefore expected that simulations performed with lower values of $p_1$ exhibit discrepancies from the theoretical predictions.

\section{Short discussion}

In this last section, we would like to discuss the implications of our theoretical predictions. 
The non-equipartition of opinion observed in MR is a striking feature that implies that the state of a social agent strongly depends on its connectivity. One expects that this effect 
should also take place in other models for opinion formation, at least if the frequency of interaction between agents is heterogeneous. Non-equipartition of opinion should be searched in empirical data, in the Blogosphere for instance, by averaging the state of agents having the same connectivity and looking for a relation between this average value and the connectivity. Such empirical study could also useful in order to test the validity of MR as a relevant model for interacting social agents. 

In this paper, we have also shown that the location of the order-disorder transition depends on the degree heterogeneity. This shift might have striking consequences in realistic situations as it implies that a change of the underlying topology might lead to a transition and to the emergence of a new (ordered or disordered) phase in the network. Similar shifts should also be searched in other models for cooperative behaviour, e.g. 
Ising models, language dynamics, etc, when they are applied to heterogeneous networks and, in particular, to  dichotomous networks. 

\acknowledgments
This work has been supported by European Commission Project 
CREEN FP6-2003-NEST-Path-012864. I would like to thank 
J. Rouchier, M. Ausloos and J.-P. Boon for fruitful discussions.


\begin{thebibliography}{0}

\bibitem{galam}
\Name{S. Galam} 
\REVIEW{Physica}{274}{1999}{132}.

\bibitem{szn}
\Name{ K. Sznajd-Weron \and J. Sznajd}
\REVIEW{Int. J. Mod. Phys. C}{11}{2000}{1157}.

\bibitem{watts2002}
\Name{D. J. Watts}
\REVIEW{Proc. Natl. Acad. Sci. USA}{99}{2002}{5766}.

\bibitem{boguna}
\Name{M. Bog\~un\'a, R. Pastor-Satorras \and A. Vespignani}
\REVIEW{Phys. Rev. Lett.}{90}{2003}{028701}.

\bibitem{dall}
\Name{L. Dall'Asta, A. Baronchelli, A. Barrat \and V. Loreto}
\REVIEW{Europhys. Lett.}{73}{2006}{969}.

\bibitem{bara}
\Name{A.-L. Barab\'asi \and R. Albert}
\REVIEW{Science}{286}{1999}{509}
   
\bibitem{sood}
\Name{V. Sood and S. Redner}
\REVIEW{Phys. Rev. Lett.}{94}{2005}{178701}.


\bibitem{lambiotte}
\Name{R. Lambiotte, M. Ausloos \and J. A. Ho{\l}yst}
\REVIEW{Phys. Rev. E}{75}{2007}{030101(R)}.

  \bibitem{lam2}
 \Name{R. Lambiotte \and M. Ausloos}
 \REVIEW{physics/0703266}{}{}{}.

\bibitem{bar}
\Name{ A. Barrat \and M. Weight}
 \REVIEW{Eur. Phys. J. B}{13}{2000}{547}.

  \bibitem{viral} 
 \Name{J. Leskovec, L. A. Adamic, B. A. Huberman}
 \REVIEW{physics/0509039}{}{}{}.
  
  \bibitem{lam0}
 \Name{R. Lambiotte, S. Thurner \and R. Hanel}
 \REVIEW{physics/0612025}{}{}{}.

\bibitem{krap}
\Name{P. L. Krapivsky \and S. Redner}
\REVIEW{Phys. Rev. Lett.}{90}{2003}{238701}.

\bibitem{liggett}
\Name{T. M. Liggett}
\Book{Interacting Particle Systems}
  \Publ{Springer-Verlag, New York}
  \Year{1985}.
  
  \bibitem{redner}
 \Name{P. Chen \and S. Redner}
  \REVIEW{Phys. Rev. E.}{71}{2005}{036101}.
 
  \bibitem{suchecki}
 \Name{K. Suchecki, V. M. Egu\'iluz \and M. San Miguel}
 \REVIEW{Phys. Rev. E}{72}{2005}{036132}.
 
 \bibitem{bipartite}
\Name{M. E. J. Newman}
\REVIEW{Proc. Natl. Acad. Sci. USA}{98}{2001}{404}.

 \bibitem{garzo5}
  \Name{V. Garzo  \and  J. Dufty} 
  \REVIEW{Phys. Rev. E}{60}{1999}{1}.
  
  \bibitem{lambi}
 \Name{R. Lambiotte \and L. Brenig}
 \REVIEW{Phys. Rev. E}{71}{2005}{042301}.
 
 \bibitem{ben1}
 \Name{E. Ben-Naim \and P. L. Krapivsky}
\REVIEW{Phys. Rev. E}{61}{2000}{R5}.

\bibitem{visone}
http://www.visone.de

 \bibitem{boguna2} 
\Name{M. Bog\~un\'a \and R. Pastor-Satorras}
 \REVIEW{Phys. Rev. E}{68}{2003}{036112}.
 
  \bibitem{fron}
\Name{A. Fronczak and P. Fronczak}
 \REVIEW{Phys. Rev. E}{74}{2006}{026121}.
 
 \bibitem{stefan} 
 \Name{S. Abe \and S. Thurner}
 \REVIEW{Phys. Rev. E}{72}{2005}{036102}. 

 \bibitem{beck} 
 \Name{C. Beck}
 \REVIEW{Phys. Rev. Lett.}{87}{2001}{180601}. 
 
  \bibitem{beck2} 
 \Name{C. Beck and E. G. D. Cohen}
 \REVIEW{Physica A}{322}{2003}{267}. 
 
 \bibitem{ausloos1}
 \Name{M. Ausloos \and R. Lambiotte}
 \REVIEW{Phys. Rev. E}{73}{2006}{11105}. 
 
\end{thebibliography}
\end{document}